# SOA Framework for Integrated Business

*Abstract* - **In this paper SOA model for business is presented. This is a survey paper. As the new technology evolve the new businesses use SOA framework.**

Close examination of SOA and EA and their corresponding governance reveal a great deal of overlap in their concepts, activities, processes, and outcomes. For example, both require input based on business objectives and produce outcomes that are tied to and measured against these objectives. Furthermore, both aim to address issues on the enterprise level (strategy and planning, reference architecture, and so on), and at the same time their governance models are similar. An enterprise that's adopting SOA while developing EA and its governances may encounter problems if the similarities and overlaps between EA and SOA are not recognized and accounted for.

The content of this series of articles is based on the practical experiences we gained while involved in a large engagement with a Fortune 500 company in the utilities industry.

Despite a tentative economic recovery, organizations today remain under relentless pressure to curtail costs and streamline operations. Many have adopted service-oriented architecture (SOA) as a way to generate sweeping efficiencies and cost savings by replacing legacy applications with modular services that can be quickly implemented and reused.

The promises of SOA are great, yet it doesn't always deliver.

The reason is simple: Most organizations lack the methodology, process, and governance necessary to design, manage, and maximize reusable services.

Blame it on the silos. In many enterprises, isolated technical teams build services for business processes that they do not understand from an end-to-end perspective. At the same time, business units create their own services in functional isolation, resulting in a torrent of point-to-point services that are not defined and not shared across the enterprise.

It's a scenario that we have seen firsthand. PricewaterhouseCoopers recently stepped in to help a global pharmaceutical firm redress an integration SOA implementation that had become overly cumbersome. In just one example of what can go wrong, the company's business process for onboarding — assigning new employees computers, network access rights, software, data access, and work groups — had ballooned to 65 interfaces. In theory, an organization should employ a single composite service for onboarding supported by multiple coarse-grained services for specific onboarding needs, since the procedure is straightforward and should vary little among divisions.

Implementing a services framework would seem to be a logical and obvious approach to managing and reusing SOA services. And it is. But companies face myriad obstacles when they attempt to tackle an initiative of this magnitude.

Implementing a services framework is daunting for most organizations because they simply don't know how to approach SOA from a business-process perspective. That's not surprising, given that a business-process approach represents a fundamental departure from the traditional methods of designing IT services.

And it's not a simple matter. Before a business process can be translated to a reusable service, organizations must meticulously analyze and model it as a generic process. Only then can the need be translated into services that can be reused across divergent business processes.

At the same time, the IT organization must establish strong governance to ensure that the services are properly designed, maintained, and reused. Governance is the linchpin of an effective SOA implementation, yet it's also a practice that many organizations recklessly disregard. They do so at great risk, since enterprises that implement SOA without governance may suffer as services proliferate wildly without a formal service definition and reuse process. In the end, no one knows how many services are in place, where they are, or what they do. The result: services are not discovered and are not reused.

The complexity of integration can be formidable. That is why it is essential to approach SOA with the proper strategy, governance, and methodology. Service definition within a services framework can help minimize the complexities of integration by taking into account data incompatibility and integration challenges. More importantly, a strategic approach to SOA ensures that an organization has the appropriately skilled resources to make integration a successful undertaking.

Another area of IT expertise in which many companies are lacking is the ability to create common data models with strong contract and policy definitions, which are essential to ensuring that services are reusable. Organizations typically take shortcuts in building the data model, failing to recognize that the services are, at their essence, nothing more than the sum of their generic information and data models.

We also have found that business units within an organization may resist a standardized service approach because they believe their needs have unique requirements that cannot be addressed by generic services. This shortsighted approach does not take into account SOA's ability to employ extensions to services that make them reusable, yet unique, by division or geographic location.

For many organizations, SOA has become the go-to strategy for maximizing IT efficiencies and cutting costs.

But when it comes to implementing a services framework for SOA, many companies simply don't know how or where to start. And given the complexity of the initiative, that's not surprising.

We believe that companies must carefully construct a SOA implementation on a strong foundation with built-in governance and a strategy that regards integration as an end-to-end business process. We call that a Centralized Services Framework (CSF).

PricewaterhouseCoopers has pioneered a CSF that fuses the functional team's business knowledge with integration Center of Excellence (COE) technical experience. Our integration COE within the Business System Integration practice has successfully designed and modeled services for a variety of global businesses based on a CSF.

Over the years, we have refined this proven methodology to deliver forward-thinking design and integration, and we have developed unique processes and tools that help facilitate a successful enterprise-wide integration using reusable services.

Experience has taught us that it's essential to tailor the CSF to a company's unique business environment and needs. So if your organization is implementing SOA for the first time or adjusting the structure of an existing system for enterprise integration, we can help you make the most of SOA.

**Service-Oriented Architecture**

The example architecture frameworks, which are presented above, reveal that the use of reusable software components becomes more and more popular. One of today's most popular architecture frameworks is the Service-Oriented Architecture (SOA) [HKG05]. SOA is seen as one of the key technologies to enable flexibility and reduce complexity in software systems. It follows the paradigm to explicitly separate an implementation from its interface. Such an interface is well-defined; that is, it is based on standards such as the Web Service Description Language (WSDL) [CCMW01, CMRW06]. Implementation and interface form together a component. In SOA, a component is referred to a service, but we prefer to use the term component. Components are independent of applications and the computing platforms on which they run. Components in a SOA can be connected without having knowledge of their technical details; they are loosely coupled. To connect components during runtime, SOA supports dynamic binding. For the message exchange between components, standardized communication protocols are used. Further, all the standards, which are used in a SOA, are extensible, meaning they are not limited to current standards and technologies. SOA distinguishes three different roles of components: component provider, component consumer, and component registry. It postulates a general protocol for interaction: A component provider registers at the component registry by submitting information about how to interact with its component. The component registry manages such information about all registered component providers and allows a component consumer to find an adequate component provider. Then, the component of the provider and the component of the consumer may bind and start interaction. A component has two kinds of interfaces: buy and sell interfaces. Buy interfaces specify which services are required by the component. In contrast, sell interfaces specify which services are provided by the component. So in terms of the component roles, in SOA, a component plays the consumer's role at the buy interfaces and at the sell interfaces it plays the provider's role. Apart from these technical paradigms services in SOA are also based on an economical paradigm. A service is comparable with a business unit. So it should create value for its environment. Therefore the two kinds of interfaces can be seen as the buy side and the sell side of the service. On the buy side, a service behaves as a service consumer or client and buys other services. On the sell side, a service behaves as the service provider and offers its service to other services. Services are operating as actors on a market place. This means, they offer their services to any consumer who needs it and they buy services from providers with the best value proposition. So both parties publish their needs and offerings at a repository, respectively.